 \documentclass[final,5p,times,twocolumn,authoryear]{elsarticle}

\usepackage{amsmath}
\usepackage{amssymb}
\usepackage{lipsum}

\journal{}

\begin{document}

\begin{frontmatter}



\title{Modelling of Economic Implications of Bias in AI-Powered Health Emergency Response Systems}


\author[first,second]{Katsiaryna Bahamazava}
\affiliation[first]{organization={Department of Mathematical Sciences G.L. Lagrange, Politecnico di Torino},
             addressline={Corso Duca degli Abruzzi, 24}, 
            city={Torino},
           postcode={10129}, 
            country={Italy}}
            \affiliation[second]{organization={iLaVita Nonprofit Foundation},
             addressline={}, 
            city={Torino - New York},
            postcode={}, 
            country={Italy - USA}}
\begin{abstract}
\indent We present a theoretical framework assessing the economic implications of bias in AI-powered emergency response systems. Integrating health economics, welfare economics, and artificial intelligence, we analyze how algorithmic bias affects resource allocation, health outcomes, and social welfare. By incorporating a bias function into health production and social welfare models, we quantify its impact on demographic groups, showing that bias leads to suboptimal resource distribution, increased costs, and welfare losses. The framework highlights efficiency-equity trade-offs and provides economic interpretations. We propose mitigation strategies, including fairness-constrained optimization, algorithmic adjustments, and policy interventions. Our findings offer insights for policymakers, emergency service providers, and technology developers, emphasizing the need for AI systems that are efficient and equitable. By addressing the economic consequences of biased AI, this study contributes to policies and technologies promoting fairness, efficiency, and social welfare in emergency response services.
\end{abstract}




\end{frontmatter}




Highlights
\begin{itemize}
    \item Developed a theoretical framework integrating economics and AI bias in emergency response.
    \item Incorporated emergency-specific variables like response time into the model.
    \item Demonstrated economic impacts of AI bias on resource allocation and health outcomes.
    \item Proposed strategies for mitigating AI bias in emergency healthcare systems.
    \item Provided policy insights to design equitable and efficient AI emergency services.

\end{itemize}

\section{Introduction}

The advent of artificial intelligence (AI) has significantly transformed various sectors, including healthcare, finance, and transportation. One critical area where AI's impact is increasingly profound is in emergency response systems. These systems are essential for 
 managing crises, dispatching emergency services, and ultimately saving lives. AI technologies have been integrated into emergency response to enhance decision-making processes, optimize resource allocation, and improve response times. By analyzing vast amounts of data in real-time, AI-powered systems can predict emergencies, prioritize incidents, and allocate resources more efficiently than traditional methods.

From an economic standpoint, the integration of AI into emergency response systems introduces complex questions related to resource allocation efficiency, cost minimization, and equity in healthcare delivery. Economic theories such as welfare economics and health economics provide a framework for analyzing how resources are distributed among different populations and the resulting impacts on social welfare \cite{ng1983welfare}. Concepts like the health production function \cite{thornton2002estimating}, which relates healthcare inputs to health outcomes, and the social welfare function, which aggregates individual utilities into a measure of societal well-being, are particularly relevant in this context. Additionally, principles of cost-benefit analysis and the efficiency-equity trade-off are crucial for understanding the economic implications of AI-driven decision-making in emergency services.

However, the integration of AI into emergency response systems is not without challenges. A growing body of research highlights the issue of algorithmic bias within AI models. Bias in AI can arise from various sources, including biased training data, flawed algorithms, or systemic inequalities reflected in the data. In the context of emergency response, such bias can lead to unequal treatment of different demographic groups or regions, resulting in disparities in response times and resource allocation. From an economic perspective, these disparities raise concerns about allocative efficiency and equity, potentially leading to suboptimal utilization of resources and welfare losses. For instance, communities with lower socio-economic status or minority populations may receive delayed or inadequate emergency services due to biased AI recommendations, exacerbating existing healthcare inequalities and increasing overall economic costs.

The presence of bias in AI-powered emergency response systems has significant economic implications. Inefficient resource utilization can lead to increased morbidity and mortality rates, which not only affect individual health outcomes but also impose substantial economic costs on society. These costs manifest not only in immediate healthcare expenses but also in long-term economic burdens due to loss of productivity, increased insurance premiums, and heightened demand for social services. Moreover, biased emergency responses can erode public trust in both AI technologies and emergency service providers, potentially leading to decreased compliance with emergency protocols and further inefficiencies. Economic theories related to externalities and public goods highlight how individual actions can have broader societal impacts, emphasizing the importance of equitable and efficient emergency response systems.

Despite the critical importance of this issue, there is a lack of comprehensive theoretical frameworks that examine the economic implications of bias in AI-powered emergency response systems. Existing studies often focus on the technical aspects of AI bias or its ethical and legal considerations but seldom integrate an economic perspective. A thorough understanding of the economic costs associated with AI bias is essential for policymakers, emergency service providers, and technology developers to make informed decisions about AI deployment, regulation, and bias mitigation strategies. By applying economic theories of resource allocation, welfare maximization, and cost-benefit analysis, we can develop models that quantify the economic impact of AI bias and identify strategies to mitigate its effects.

The contributions of this paper are threefold:

\begin{enumerate}
    \item \textbf{Theoretical Integration}: This work brings together concepts from economics—particularly health economics and welfare economics—artificial intelligence, and public policy to create a comprehensive framework for understanding the economic implications of AI bias in emergency response systems.

    \item \textbf{Economic Modeling}: By incorporating economic theories such as the health production function, social welfare functions, and cost-benefit analysis, the paper develops advanced economic models to quantify the costs associated with AI bias and to analyze efficiency and equity implications.

    \item \textbf{Policy and Ethical Insights}: By analyzing the economic costs alongside ethical and policy considerations, the framework provides valuable guidance for stakeholders in developing regulations and standards that promote fairness, efficiency, and social welfare maximization.

\end{enumerate}

The remainder of this paper is organized as follows. Section~\ref{sec:Literature-Review} provides a literature review on AI in emergency services, algorithmic bias, economic implications of biased systems, and relevant policy frameworks. Section~\ref{sec:theoretical-framework} details the development of the theoretical model, including the conceptual model, economic modeling, and analysis of bias mechanisms in AI from an economic perspective. Section~\ref{sec:Discussion} provides a comprehensive discussion on the implications for stakeholders, limitations of the study, and suggestions for future research. Section~\ref{sec:Conclusion} concludes the paper by summarizing the key findings and emphasizing the importance of addressing AI bias in emergency response systems to enhance economic efficiency and equity.

\section{Literature Review}
\label{sec:Literature-Review}
The integration of artificial intelligence (AI) into emergency management and healthcare has garnered significant attention due to its potential to enhance crisis response, resource allocation, and decision-making processes. However, the deployment of AI technologies is accompanied by ethical, operational, and economic challenges, particularly concerning bias and fairness. Addressing these challenges is crucial to optimize AI's impact on emergency systems and mitigate potential economic and societal costs.

AI technologies have been increasingly applied in emergency healthcare settings to improve efficiency and patient outcomes. Visave (2024) discusses the transformative potential of AI in emergency management, highlighting improvements in response times and resource distribution. Despite these advantages, the author points out that AI systems often perpetuate biases against marginalized groups due to skewed training data, resulting in unfair resource allocation and exacerbating existing inequalities. The opacity of AI models, often referred to as "black boxes," further complicates efforts to ensure transparency and accountability. Visave emphasizes the necessity of ethical AI deployment, underscoring data privacy concerns and the risk of undermining public trust if biases are not effectively addressed \cite{visave2024ai}.

Building on these concerns, Pasli et al. (2024) explore the application of AI in emergency department (ED) triage. Their study evaluates the use of AI, specifically ChatGPT, in predicting triage outcomes and finds that AI can achieve near-perfect agreement with human triage teams. However, they caution that reliance on AI must be carefully regulated to avoid ethical pitfalls, such as breaches in data privacy and over-reliance on AI in medical decision-making. The study reinforces the need for maintaining human oversight to ensure responsible integration of AI into healthcare systems, highlighting that technical accuracy alone is insufficient without ethical considerations \cite{pasli2024assessing}.

Piliuk et al. (2023) provide a broader review by categorizing AI applications in emergency medicine into diagnostics and triage. AI has demonstrated significant utility in disease prediction and prioritizing patients based on the severity of their conditions. Despite these benefits, the authors highlight key challenges such as the lack of generalization across diseases and the need for more explainable AI models. These challenges are central to the discourse on AI bias, where models performing well in controlled environments may struggle in diverse real-world scenarios, potentially leading to disparities in healthcare outcomes \cite{piliuk2023artificial}.

Cheng et al. (2024) delve into the specific barriers and facilitators to implementing AI in emergency departments. They identify challenges such as model interpretability, physician autonomy, and medicolegal concerns as significant barriers to adoption. Conversely, aligning AI systems with existing clinical processes and providing adequate training to end-users are key facilitators. The authors emphasize that successful integration of AI in emergency settings depends not only on technical accuracy but also on addressing broader ethical, legal, and social concerns to ensure transparency and trustworthiness \cite{cheng2024implementation}.

Bias in AI systems poses significant risks to equitable healthcare delivery. Focusing on medical imaging, Jones et al. (2024) offer a causal framework for understanding various forms of bias—prevalence disparities, presentation disparities, and annotation disparities. Each form requires different mitigation strategies. By adopting a causal perspective, the authors provide a nuanced understanding of how biases emerge and affect the fairness of AI systems in medical contexts. They call for greater transparency in data collection and emphasize that no single bias mitigation technique can comprehensively address all forms of bias, reinforcing the need for tailored solutions across different AI applications \cite{jones2024causal}.

Similarly, Flores et al. (2024) examine biases in AI systems used for public health surveillance. They argue that non-representative data can lead to misrepresentation of certain populations, particularly those from lower socio-economic backgrounds, thereby exacerbating health disparities. The paper advocates for open collaboration among stakeholders, fairness auditing, and universal guidelines to mitigate these risks. Such measures are essential to ensure that AI systems are not only technically robust but also equitable, especially when influencing public health policy \cite{flores2024addressing}.

The economic consequences of bias in healthcare systems are profound. Mackenbach et al. (2011) analyze the substantial economic burden caused by socioeconomic inequalities in health across the European Union (EU). They estimate that inequality-related health losses lead to over 700,000 deaths and 33 million cases of ill health annually. These disparities account for 20\% of healthcare costs and 15\% of social security benefits in the EU, reducing labor productivity and resulting in a 1.4\% decrease in the EU's Gross Domestic Product (GDP) annually. The total welfare loss due to health inequalities is estimated at €980 billion per year, or 9.4\% of GDP. The findings underscore the need for investments in policies aimed at reducing health inequalities, which could yield significant economic benefits alongside improved health outcomes \cite{mackenbach2011economic}.

Similarly, Pickwell-Smith et al. (2024) conduct a systematic review to explore socioeconomic disparities in colorectal cancer care within the UK. They identify significant inequalities in survival rates between patients from deprived areas and those from affluent regions. Socioeconomic factors influence the timeliness of diagnosis, access to surgery, and the receipt of chemotherapy and radiotherapy. The study highlights that patients from deprived backgrounds experience delays in treatment and are less likely to receive life-saving interventions, despite the UK's universal healthcare system. The authors call for further research to understand the underlying causes of these inequalities and recommend policies to reduce barriers to effective cancer care, emphasizing the broader economic and social implications of bias in healthcare systems \cite{pickwell2024inequalities}.

Wolff et al. (2020) provide a comprehensive review of the cost-effectiveness of AI in healthcare. They highlight a significant lack of robust economic impact assessments, noting that most studies focus on specific cost-saving aspects rather than providing complete cost-benefit analyses. Key gaps include the failure to account for initial investment and operational costs of AI infrastructure and a lack of comparison with alternative technologies. The authors call for more comprehensive studies that include detailed cost-effectiveness analyses to support decision-making in AI implementation. This is particularly relevant for assessing the economic implications of bias in AI-powered emergency response systems, where similar economic assessment gaps exist \cite{wolff2020economic}.

Extending beyond healthcare, Mohla et al. (2024) explore the gendered effects of AI in the labor market. They argue that while efforts to reduce algorithmic bias are important, AI's broader economic and structural effects—such as automating tasks that disproportionately affect women—need greater attention. These changes exacerbate existing inequalities and increase job insecurity for marginalized groups. This perspective highlights the far-reaching implications of AI bias beyond specific technological applications, emphasizing the importance of considering societal structures in AI deployment \cite{mohla2024thinking}.

In the context of financial inclusion, Adeoye (2024) investigates the role of AI in developing economies. AI-driven tools like machine learning models for credit scoring can help underserved populations access financial services. However, challenges such as ensuring fairness, transparency, data privacy concerns, and the digital divide limit access to AI-driven services. These considerations align with the broader need to develop AI systems that are both efficient and equitable to prevent unintended economic consequences \cite{adeoye2024leveraging}.

The reviewed literature underscores the dual potential and challenges of integrating AI into emergency management and healthcare systems. While AI offers significant benefits in improving efficiency, accuracy, and resource allocation, there are pervasive concerns about algorithmic bias leading to inequities in service delivery. These biases not only raise ethical and social justice issues but also have substantial economic implications, including increased healthcare costs, reduced productivity, and welfare losses. The lack of comprehensive economic evaluations, as highlighted by Wolff et al. (2020), indicates a gap in understanding the full impact of AI implementation in healthcare settings.

Furthermore, studies such as those by Mackenbach et al. (2011) and Pickwell-Smith et al. (2024) illustrate the profound economic burden of health inequalities, emphasizing the need for policies that address both the ethical and economic dimensions of healthcare disparities. The existing literature calls for multidisciplinary approaches combining technical, ethical, and economic analyses to develop AI systems that are both effective and equitable.

This paper seeks to address the identified gaps by developing a theoretical framework that integrates economic theories with AI bias considerations in emergency response systems. By examining the economic implications of algorithmic bias, the study aims to provide valuable insights for policymakers, emergency service providers, and technology developers. The framework will contribute to the broader discourse on responsible AI deployment, emphasizing the importance of economic evaluations in understanding and mitigating the ramifications of biased AI systems.
\section{Theoretical Framework Development}
\label{sec:theoretical-framework}
In this section, we develop a comprehensive theoretical framework to assess the economic implications of bias specifically arising from AI and machine learning (ML) systems in emergency response. Recognizing the complexity of healthcare economics and the unique dynamics of emergency medicine, we integrate advanced economic models with AI bias considerations, emphasizing emergency-specific variables. Our framework is grounded in welfare economics, health production functions, resource allocation theories, and incorporates the mechanics of AI/ML systems within the context of emergency medicine to provide a robust analysis suitable for health economics research.

\subsection{Assumptions and Justifications}

Before delving into the model, we explicitly state and justify the underlying assumptions, emphasizing how bias is related to AI and ML systems in emergency medicine:

\begin{enumerate}
    \item \textbf{AI-Driven Emergency Resource Allocation}: Emergency resources, such as ambulances, emergency medical teams, and critical care units, are allocated based on decisions made by AI/ML systems. These systems use algorithms trained on historical emergency data to predict incident locations, severity, and allocate resources accordingly. This reflects the real-world shift towards AI-assisted decision-making in emergency services.

    \item \textbf{Bias Originating from AI/ML Systems in Emergency Contexts}: Bias in the AI system arises due to factors such as biased training data, algorithmic design, and systemic inequalities reflected in emergency response records. We represent this bias using a multiplicative bias function \( B(g_i) \geq 1 \) specific to each group \( g_i \). A value of \( B(g_i) = 1 \) indicates no bias from the AI/ML system, while \( B(g_i) > 1 \) reflects the degree of bias introduced against group \( g_i \) in emergency resource allocation.

    \item \textbf{Resource Scarcity and Time Sensitivity}: Emergency resources are limited, and their timely allocation is critical. This aligns with the economic principle of scarcity and reflects real-world constraints in emergency services, where delays can significantly impact health outcomes.

    \item \textbf{Health Production Function with Time Sensitivity}: The health outcomes for each group are determined by a health production function \( H(g_i) = h(RA(g_i), RT(g_i)) \), where \( RA(g_i) \) is the amount of emergency resources allocated to group \( g_i \) and \( RT(g_i) \) is the response time for group \( g_i \). The function exhibits diminishing marginal returns with respect to resources (\( \frac{\partial h}{\partial RA} > 0, \frac{\partial^2 h}{\partial RA^2} < 0 \)) and a negative impact with respect to response time (\( \frac{\partial h}{\partial RT} < 0 \)).

    \item \textbf{Utility Dependence on Health Outcomes}: The utility of each group depends solely on their health outcomes, \( U(g_i) = u(H(g_i)) \), with \( u' > 0 \). This aligns with utility theory in health economics, where better health outcomes increase individual and societal utility.

    \item \textbf{Emergency Incident Rates and Severity}: Each group \( g_i \) has a specific emergency incident rate \( EIR(g_i) \) and severity \( S(g_i) \), influencing the optimal allocation of resources. The AI/ML system aims to predict these factors to allocate resources effectively.

    \item \textbf{Social Welfare Maximization}: The objective is to maximize a social welfare function \( W \) that aggregates individual utilities, considering both efficiency and equity in emergency healthcare delivery.

    \item \textbf{Cost Structure in Emergency Contexts}: The total economic cost comprises direct costs of resource allocation, costs associated with response times (e.g., delays), and indirect costs associated with health outcomes resulting from emergencies.

\end{enumerate}

\subsection{Incorporating Emergency-Specific Variables into the Model}

\subsubsection{Identification of Key Variables}

We define the following variables and functions specific to emergency medicine:

\begin{itemize}
    \item \textbf{Population Groups} \( G = \{ g_1, g_2, \dots, g_n \} \): Different demographic or regional groups within the population.

    \item \textbf{Bias Function} \( B(g_i) = f_{\text{bias}}(D(g_i), A(g_i)) \): Represents the level of AI/ML-induced bias against group \( g_i \) in emergency resource allocation, influenced by data quality \( D(g_i) \) and algorithmic fairness \( A(g_i) \).

    \item \textbf{Data Quality Function} \( D(g_i) \): Measures the adequacy and representativeness of training data for group \( g_i \). Higher values indicate better data quality (less data bias).

    \item \textbf{Algorithmic Fairness Function} \( A(g_i) \): Quantifies the fairness of the AI/ML algorithm's treatment of group \( g_i \). Higher values indicate greater algorithmic fairness.

    \item \textbf{Emergency Resource Allocation} \( RA(g_i) \): The amount of emergency resources allocated to group \( g_i \).

    \item \textbf{Optimal Emergency Resource Allocation} \( RA^*(g_i) \): The ideal allocation without bias, determined by emergency incident rates \( EIR(g_i) \) and severity \( S(g_i) \).

    \item \textbf{Response Time} \( RT(g_i) \): The time it takes for emergency resources to reach group \( g_i \).

    \item \textbf{Optimal Response Time} \( RT^*(g_i) \): The minimal response time achievable without bias.

    \item \textbf{Health Outcome} \( H(g_i) = h(RA(g_i), RT(g_i)) \): The morbidity and mortality outcomes for group \( g_i \), dependent on allocated resources and response time.

    \item \textbf{Utility Function} \( U(g_i) = u(H(g_i)) \): The utility derived by group \( g_i \) from their health outcomes.

    \item \textbf{Emergency Incident Rate} \( EIR(g_i) \): The rate at which emergencies occur in group \( g_i \).

    \item \textbf{Severity of Emergencies} \( S(g_i) \): The average severity level of emergencies experienced by group \( g_i \).

    \item \textbf{Social Welfare Function} \( W = \sum_{i=1}^{n} w_i U(g_i) \): An aggregate measure of societal welfare based on the utilities of all groups.

    \item \textbf{Total Economic Cost} \( C_{\text{total}} \): The sum of direct costs of resource allocation, costs associated with response times, and indirect costs due to health outcomes.
\end{itemize}

To improve readability and ensure consistent notation, we provide a summary table of all variables and their definitions (see Table~\ref{tab:variables}).

\begin{table}[htp]
\centering
\caption{Summary of Variables and Functions}
\small 
\renewcommand{\arraystretch}{1.1} 
\setlength{\tabcolsep}{4pt} 
\begin{tabular}{p{0.2\linewidth} p{0.7\linewidth}} 
\hline
\textbf{Variable} & \textbf{Definition} \\
\hline
\( G  \{ g_1, g_2, \dots, g_n \} \) & Set of population groups \\
\( B(g_i) \) & Bias function for group \( g_i \) \\
\( D(g_i) \) & Data quality function for group \( g_i \) \\
\( A(g_i) \) & Algorithmic fairness function for group \( g_i \) \\
\( RA(g_i) \) & Emergency resource allocation to group \( g_i \) \\
\( RA^*(g_i) \) & Optimal emergency resource allocation for group \( g_i \) \\
\( RT(g_i) \) & Response time for group \( g_i \) \\
\( RT^*(g_i) \) & Optimal response time for group \( g_i \) \\
\( H(g_i) \) & Health outcome for group \( g_i \) \\
\( h(\cdot) \) & Health production function \\
\( U(g_i) \) & Utility of group \( g_i \) \\
\( u(\cdot) \) & Utility function \\
\( EIR(g_i) \) & Emergency incident rate for group \( g_i \) \\
\( S(g_i) \) & Severity of emergencies for group \( g_i \) \\
\( W \) & Social welfare function \\
\( w_i \) & Weight assigned to group \( g_i \) in social welfare \\
\( C_{\text{total}} \) & Total economic cost \\
\( C_D(D(g_i)) \) & Cost of improving data quality for group \( g_i \) \\
\( C_A(A(g_i)) \) & Cost of enhancing algorithmic fairness for group \( g_i \) \\
\( C_{\text{bias\_reduction}} \) & Total cost of reducing AI/ML bias \\
\( \lambda \) & Marginal utility of income \\
\hline
\end{tabular}
\label{tab:variables}
\end{table}

\subsubsection{Establishing Relationships}

\paragraph{Bias Function Derived from AI/ML Characteristics}

We model the bias function \( B(g_i) \) as:
\begin{equation}
B(g_i) = \frac{1}{D(g_i) \cdot A(g_i)}
\end{equation}
Where:

\begin{itemize}
    \item \( D(g_i) \in (0,1] \): Higher values indicate better data quality (less data bias).
    \item \( A(g_i) \in (0,1] \): Higher values indicate greater algorithmic fairness.
\end{itemize}

A lower \( D(g_i) \) or \( A(g_i) \) increases \( B(g_i) \), indicating more bias against group \( g_i \).

\paragraph{Emergency Resource Allocation Influenced by AI/ML Bias}

The AI/ML system allocates resources based on its predictions, influenced by bias:
\begin{equation}
RA(g_i) = RA^*(g_i) \cdot D(g_i) \cdot A(g_i)
\end{equation}

This reflects that better data quality and algorithmic fairness lead to allocations closer to the optimal \( RA^*(g_i) \).

\paragraph{Response Time Influenced by AI/ML Bias}

Bias can also affect response times due to misallocation of resources:

\begin{equation}
RT(g_i) = \frac{RT^*(g_i)}{D(g_i) \cdot A(g_i)} = RT^*(g_i) \cdot B(g_i)
\end{equation}

Where \( RT^*(g_i) \) is the optimal (minimal) response time without bias.

\paragraph{Health Outcomes in Emergency Contexts}

The health production function incorporates both resource allocation and response time:
\begin{equation}
H(g_i) = h(RA(g_i), RT(g_i))
\end{equation}

With:

\[
\frac{\partial h}{\partial RA} > 0, \quad \frac{\partial^2 h}{\partial RA^2} < 0, \quad \frac{\partial h}{\partial RT} < 0
\]

This reflects that increased resources and shorter response times improve health outcomes.

\paragraph{Utility and Health Outcomes}

Utility depends on health outcomes:

\begin{equation}
U(g_i) = u(H(g_i))
\end{equation}

With \( u' > 0 \).

\paragraph{Social Welfare Function}

Social welfare aggregates utilities:

\begin{equation}
W = \sum_{i=1}^{n} w_i U(g_i)
\end{equation}

\paragraph{Economic Cost Function}

Total economic cost includes:

\begin{equation}
C_{\text{total}} = \sum_{i=1}^{n} \left[ C_{\text{resource}}(RA(g_i)) + C_{\text{response}}(RT(g_i)) + C_{\text{health}}(H(g_i)) \right]
\end{equation}

\subsubsection{Optimization Problem in Emergency Context}

The objective function now explicitly involves AI/ML system characteristics and emergency-specific variables:

\begin{equation}
\begin{aligned}
\max_{\{ D(g_i), A(g_i) \}} \quad & W - \lambda \left( C_{\text{total}} + C_{\text{bias\_reduction}} \right) \\
\text{subject to} \quad & RA(g_i) = RA^*(g_i) \cdot D(g_i) \cdot A(g_i), \quad \forall i \\
& RT(g_i) = \frac{RT^*(g_i)}{D(g_i) \cdot A(g_i)}, \quad \forall i \\
& D(g_i), A(g_i) \in (0,1], \quad \forall i \\
& \sum_{i=1}^{n} RA(g_i) \leq RA_{\text{total}} \\
\end{aligned}
\end{equation}

Where \( C_{\text{bias\_reduction}} = \sum_{i=1}^{n} \left[ C_D(D(g_i)) + C_A(A(g_i)) \right] \) represents the costs of improving data quality and algorithmic fairness, and \( \lambda \) is the marginal utility of income.

\subsection{Economic Interpretation of Mathematical Results in Emergency Contexts}

\subsubsection{Impact of AI/ML Bias on Health Outcomes}

Bias in AI/ML systems (\( D(g_i) < 1 \), \( A(g_i) < 1 \)) reduces \( RA(g_i) \) and increases \( RT(g_i) \), leading to lower \( H(g_i) \) and \( U(g_i) \), and thus decreasing \( W \). The time-sensitive nature of emergency interventions means that even small increases in response time can have significant adverse effects on morbidity and mortality.

\subsubsection{Marginal Benefits of Reducing AI/ML Bias}

The marginal benefit of improving data quality or algorithmic fairness for group \( g_i \) is:
\begin{multline}
\frac{\partial W}{\partial D(g_i)} = 
    \frac{\partial W}{\partial U(g_i)} \cdot 
    \frac{\partial U(g_i)}{\partial H(g_i)} \cdot \left( 
    \frac{\partial H(g_i)}{\partial RA(g_i)} \cdot RA^*(g_i) \cdot A(g_i) 
    \right. \\
    \left. - \frac{\partial H(g_i)}{\partial RT(g_i)} \cdot 
    \frac{RT^*(g_i)}{(D(g_i) \cdot A(g_i))^2} \cdot A(g_i) \right)
\end{multline}

Similarly for \( \frac{\partial W}{\partial A(g_i)} \). This expression shows how investments in data quality and algorithmic fairness can yield welfare gains by improving resource allocation and reducing response times.

\subsubsection{Cost of Reducing AI/ML Bias}

Improving \( D(g_i) \) and \( A(g_i) \) involves costs \( C_D(D(g_i)) \) and \( C_A(A(g_i)) \), respectively. These costs must be weighed against the marginal benefits in terms of improved health outcomes and social welfare.

\subsection{Unique Challenges in Emergency Medicine and AI Bias}

Emergency medicine is characterized by unpredictability, urgency of care, and the critical impact of delays. AI/ML bias in this context can have immediate and severe consequences:

\begin{itemize}
    \item \textbf{Unpredictability of Emergencies}: Biases in AI predictions can misrepresent the likelihood or location of emergencies, leading to resource misallocation.

    \item \textbf{Urgency of Care}: Delays caused by biased response times can result in preventable deaths or long-term health complications.

    \item \textbf{Critical Impact of Delays}: In emergency medicine, every minute counts. AI bias that increases response times can exponentially worsen health outcomes.

    \item \textbf{Severity Considerations}: AI bias may underestimate the severity \( S(g_i) \) of emergencies in certain groups, leading to insufficient resource allocation.
\end{itemize}

\subsection{Strategies for Bias Mitigation in Emergency AI/ML Systems}

\subsubsection{Improving Data Quality in Emergency Contexts}

\begin{itemize}
    \item Collecting comprehensive emergency incident data across all groups.
    \item Addressing underreporting and inaccuracies in emergency response records.
    \item Incorporating real-time data to adjust for dynamic emergency situations.
\end{itemize}

\subsubsection{Enhancing Algorithmic Fairness with Emergency-Specific Considerations}

\begin{itemize}
    \item Incorporating fairness constraints that account for emergency incident rates and severity.
    \item Utilizing algorithms designed to prioritize response time minimization across all groups.
    \item Implementing robust validation using diverse emergency scenarios to ensure equitable performance.
\end{itemize}

\subsubsection{Policy Interventions Specific to Emergency Services}

\begin{itemize}
    \item \textbf{Regulations on Emergency AI Fairness}: Mandating audits that specifically evaluate AI performance in emergency response times and resource allocation equity.

    \item \textbf{Transparency and Explainability}: Requiring AI systems to provide interpretable predictions and justifications for resource allocation decisions in emergencies.

    \item \textbf{Emergency Preparedness Funding}: Allocating resources to improve infrastructure and reduce response times, particularly in underserved areas.
\end{itemize}

\subsection{Implications and Extensions}

By incorporating emergency-specific variables and adjusting the health production function to reflect the time-sensitive nature of emergency interventions, our framework provides a more accurate representation of the economic implications of AI bias in emergency medicine. The model underscores the critical importance of minimizing response times and ensuring equitable resource allocation.

Future research can:

\begin{itemize}
    \item Empirically estimate the impact of AI bias on response times and health outcomes in emergency settings.
    \item Develop dynamic models that account for the stochastic nature of emergencies.
    \item Explore the integration of behavioral economics to consider how human factors influence emergency response and AI system interactions.
\end{itemize}

By addressing the unique challenges in emergency medicine, our framework enhances the understanding of how AI biases can have immediate and severe consequences, guiding the development of AI systems that are both efficient and equitable in emergency healthcare delivery.

\section{Discussion}
\label{sec:Discussion}
In this section, we discuss the implications of our theoretical model for various stakeholders, outline the limitations of our study, and suggest directions for future research.

\subsection{Implications for Stakeholders}

\subsubsection{Policymakers}

Our model highlights the significant economic costs associated with bias in AI-powered emergency response systems. For policymakers, this underscores the importance of implementing regulations and guidelines that enforce fairness and equity in AI algorithms. Policies could include mandatory bias audits, transparency requirements, and incentives for equitable resource allocation. By establishing clear standards and accountability measures, policymakers can help ensure that AI technologies contribute positively to public health outcomes without exacerbating existing disparities.

\subsubsection{Emergency Service Providers}

Emergency service providers stand to benefit from adopting AI systems that are free from bias, as this can improve the efficiency of resource allocation and enhance health outcomes. Our framework suggests that investing in bias mitigation strategies can lead to better utilization of resources and reduce overall costs associated with emergency responses. Providers should prioritize training and awareness programs to understand how AI bias can impact their operations and should collaborate with technology developers to implement solutions that align with ethical standards.

\subsubsection{Technology Developers}

For technology developers, our study emphasizes the need to design AI algorithms with built-in fairness constraints and to regularly assess and adjust models to minimize bias. Incorporating ethical considerations into the development process can improve the performance of AI systems and ensure compliance with regulatory standards. Developers should engage in interdisciplinary collaboration with healthcare professionals and ethicists to create algorithms that are both effective and equitable.

\subsubsection{Patients and the Public}

The general public, especially marginalized communities, are directly affected by biases in emergency response systems. Addressing these biases can lead to improved health outcomes and increased trust in AI technologies. Public awareness campaigns and community engagement initiatives can empower individuals to advocate for fairer AI systems. Transparency in how AI decisions are made can also help build public confidence.

\subsection{Limitations of the Study}

While our theoretical framework provides valuable insights, there are limitations to consider:

\begin{itemize}
    \item \textbf{Theoretical Assumptions}: The model is based on theoretical constructs and assumptions that may not fully capture the complexities of real-world emergency response systems. Factors such as human behavior, unforeseen emergencies, and infrastructure variability are difficult to model precisely.
    \item \textbf{Lack of Empirical Validation}: The absence of empirical data limits the ability to validate the model's predictions. Real-world data on resource allocation, bias levels, and economic costs are needed to test and refine the theoretical constructs.
    \item \textbf{Homogeneity Assumption}: The model assumes a homogeneous response to resource allocation across different groups, which may not account for varying social determinants of health, cultural differences, or regional disparities.
    \item \textbf{Dynamic Factors}: Emergency response systems are dynamic, with factors such as time-sensitive decision-making and evolving emergencies. The static nature of the model may not fully reflect these dynamics.
\end{itemize}

\subsection{Suggestions for Future Research}

Future research should focus on the following areas:

\begin{itemize}
    \item \textbf{Empirical Validation}: Collecting and analyzing real-world data on emergency response times, resource allocation, health outcomes, and economic costs across different demographic groups to validate and refine the theoretical model.
    \item \textbf{Model Enhancement}: Extending the model to incorporate additional variables such as social determinants of health, varying behavioral responses, and infrastructure constraints to enhance its applicability and accuracy.
    \item \textbf{Algorithm Development}: Developing and testing AI algorithms with integrated bias mitigation techniques in real-world settings to assess their effectiveness and impact on economic outcomes.
    \item \textbf{Interdisciplinary Collaboration}: Encouraging collaboration among economists, data scientists, healthcare professionals, ethicists, and policymakers to develop comprehensive solutions that address both technical and societal aspects of AI bias.
    \item \textbf{Policy Impact Studies}: Investigating the effectiveness of different policy interventions and regulatory frameworks in reducing AI bias and improving economic and health outcomes.
    \item \textbf{Public Engagement Research}: Exploring strategies for increasing public awareness and engagement regarding AI bias in emergency services to foster community-driven demand for equitable technologies.
\end{itemize}

By addressing these areas, future studies can build upon our theoretical framework to develop practical solutions that mitigate bias in AI-powered emergency response systems, ultimately leading to improved health outcomes and economic efficiency.

\section{Conclusion}
\label{sec:Conclusion}
As AI technologies become increasingly integral to critical public services, particularly in emergency response systems, understanding and addressing the implications of algorithmic bias is paramount. This paper has shed light on the significant economic consequences of biased AI in emergency response systems, demonstrating how such biases lead to suboptimal resource allocation, increased economic costs, and welfare losses. By developing a comprehensive theoretical framework that integrates advanced economic modeling with AI bias considerations, we provide stakeholders with the tools necessary to make informed decisions that enhance both the fairness and efficiency of emergency services.

Our framework bridges the gap between economic theory and the technical aspects of AI, explicitly modeling how biases in AI and machine learning systems affect resource allocation, health outcomes, and social welfare. By incorporating mechanisms of AI-induced bias—such as data quality issues and algorithmic fairness—we offer a nuanced understanding of the economic implications of biased AI-powered emergency response systems. This interdisciplinary approach not only contributes to the broader discourse on responsible AI deployment but also provides practical strategies for bias mitigation, including improving data collection, enhancing algorithmic design, and implementing policy interventions.

The findings of this study have significant implications for policymakers, healthcare providers, and technology developers. Policymakers are equipped with a framework that highlights the trade-offs between efficiency and equity, guiding the development of regulations and standards that promote fairness and transparency in AI algorithms. 
 Healthcare providers can utilize the insights to optimize resource allocation and improve health outcomes, while technology developers are encouraged to design AI systems that are both effective and equitable.

Looking forward, the potential of our framework to inform future research and policy development is substantial. By providing a robust foundation for empirical validation and further theoretical exploration, our model invites researchers to expand upon its 
 constructs, incorporate dynamic and stochastic factors, and apply it to various contexts within healthcare and other sectors affected by AI bias. Policymakers can leverage the insights from our framework to craft informed regulations and standards that address the economic and ethical challenges posed by AI bias. Ultimately, we envision our framework serving as a catalyst for interdisciplinary collaboration and innovation, driving the creation of AI-powered systems that serve all communities equitably and effectively.

By integrating advanced economic modeling with AI bias considerations, we have provided a robust theoretical framework suitable for evaluating the economic implications of bias in 
 emergency response systems. This framework not only enhances our understanding of the complex interplay between AI technology and economic outcomes but also sets the stage for 
 future research and policy initiatives aimed at promoting fairness, efficiency, and social welfare maximization in emergency healthcare services.







\end{document}